\documentclass[preprint,aps,pre,showpacs,superscriptaddress]{revtex4}

\def \beq{\begin{equation}}         \def \eeq{\end{equation}}
\def \beqa{\begin{eqnarray}}        \def \eeqa{\end{eqnarray}}
\def \bea{\begin{array}}        \def \eea{\end{array}}

\def\bio#1#2#3{{Biophys. J. }{\bf #1}, #2 (#3)}

\def\nats#1#2#3{{Nature struct. Biol. }{\bf #1}, #2 (#3)}
\def\pnas#1#2#3{{Proc. Natl. Acad. Sci. USA }{\bf #1}, #2 (#3)}
\def\pre#1#2#3{{Phys. Rev. E }{\bf #1}, #2 (#3)}

\def\prl#1#2#3{{Phys. Rev. Lett. }{\bf #1}, #2 (#3)}
\def\sci#1#2#3{{Science }{\bf #1}, #2 (#3)}

\usepackage{amsmath}
\usepackage{epsfig}
\begin{document}

\title{Stretching single RNAs: exact numerical and stochastic simulation methods}
\author{Fei Liu}
\email[Email address:]{liufei@tsinghua.edu.cn} \affiliation{Center
for Advanced Study, Tsinghua University, Beijing 100084, China}
\author{Bi-hui Zhu}
\affiliation{Department of physics, Peking university, Beijing
100871, China}
\author{ Zhong-can Ou-Yang}
\affiliation{Center for Advanced Study, Tsinghua University,
Beijing 100084, China}\affiliation{Institute of Theoretical
Physics, The Chinese Academy of Sciences, P. O. Box 2735, Beijing
100080, China}
\date{\today}

\begin{abstract}
Exact numerical methods and stochastic simulation methods are
developed to study the force stretching single RNA issue on the
secondary structure level in equilibrium. By computing the
force-extension curves on the constant force and the constant
extension ensembles, we find the two independent methods agree
with each other quite well. To show the precision of our methods
in predicting unfolding experiments, the unfolding forces of
different RNA molecules under different experimental conditions
are calculated. We find that the ionic corrections on the RNA free
energies alone might not account for the apparent differences
between the theoretical calculations and the experimental data; an
ionic correction to the persistent length of single-stranded RNA
should be necessary.
\end{abstract}

 \maketitle
\section{introduction}
\label{introduction} In the past few years, enormous theoretical
efforts have been devoted to understanding folding/unfolding
phenomena of proteins and RNA observed in single-molecule
experiments. Diverse methods including molecular dynamics
\cite{Lu,Bryant}, Monte Carlo method \cite{Kilmov,Socci,harlepp},
and other theoretical models \cite{Hummer} have been developed.
However, these studies mainly focused on the dynamical behavior of
proteins under force, and few concerned about RNA \cite{harlepp}.
To fill this gap, we recently developed kinetic Monte Carlo
simulation methods to investigate the RNA kinetic behaviors in
constant force and constant extension ensembles on secondary
structure level \cite{liuf1,liuf2}. In addition to the intriguing
nonequilibrium phenomena, the most direct application of our
simulation methods is to investigate the relative simple unfolding
behaviors in equilibrium \cite{liuf1,liuf2}. Different from
complex protein unfolding behaviors even in equilibrium
state\cite{Shen}, force unfolding RNA has been showed to be solved
in an exact numerical way in the constant extension ensemble
\cite{gerland}, though the extension to the constant force
ensemble were not reported till now. One of natural question
arises whether our simulations are accurate enough comparing to
the numerical method. In this report, we address this question.

The organization of this paper is as follows. We first ,In Sec.
\ref{method} simply review the Monte Carlo methods developed by
us. Then the exact numerical methods for the constant force and
the constant extension ensembles are showed. In Sec. \ref{result}
we compare the simulation and numerical methods in the two
ensembles. We particularly point out the importance of persistent
length of RNA in predicting unfolding forces. Finally Sec.
\ref{conclusion} presents our conclusion.

\section{The Model and methods}
\label{method}
\begin{figure}[htpb]
\begin{center}
\includegraphics[width=0.7\columnwidth]{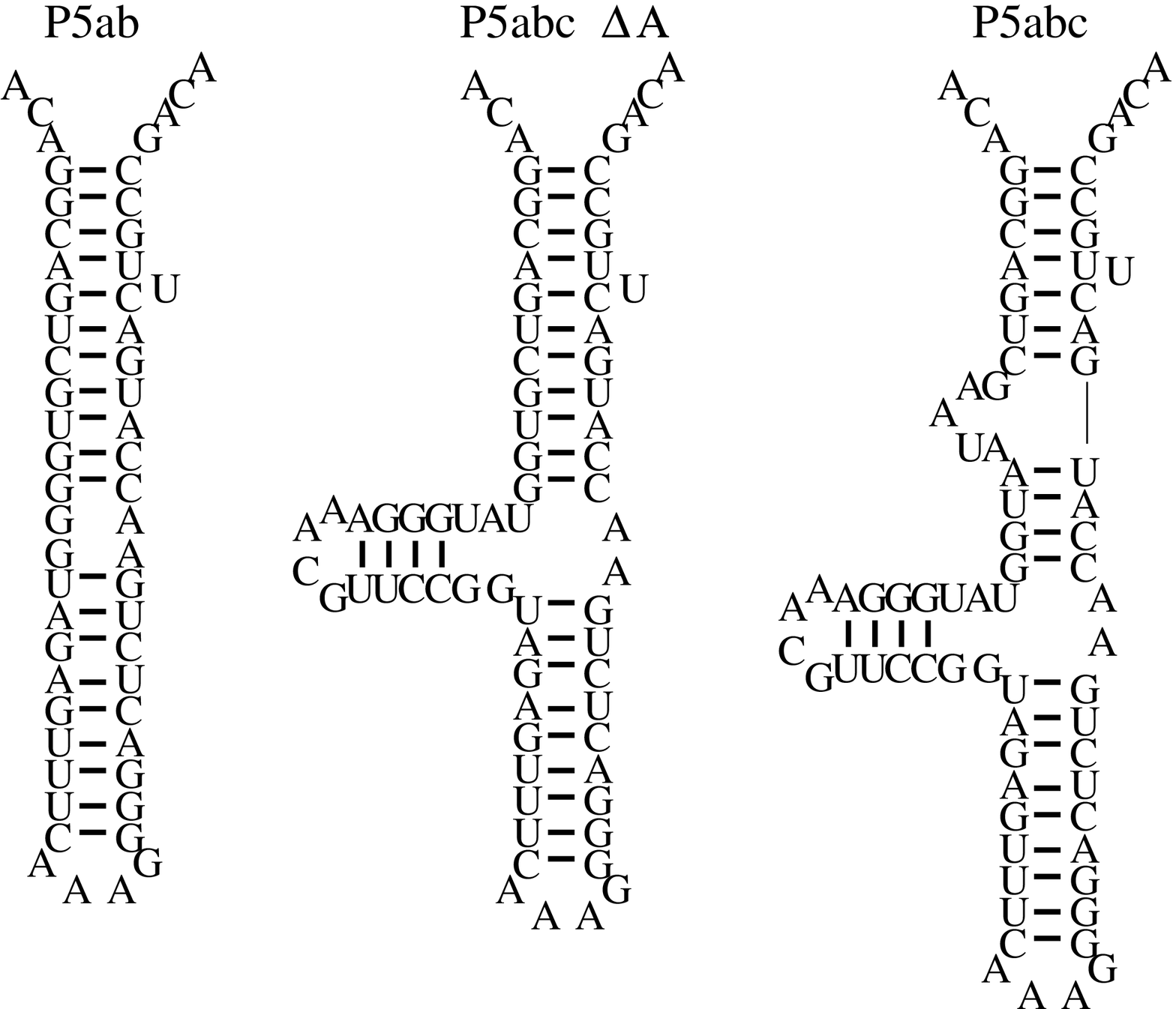}
\includegraphics[width=0.7\columnwidth]{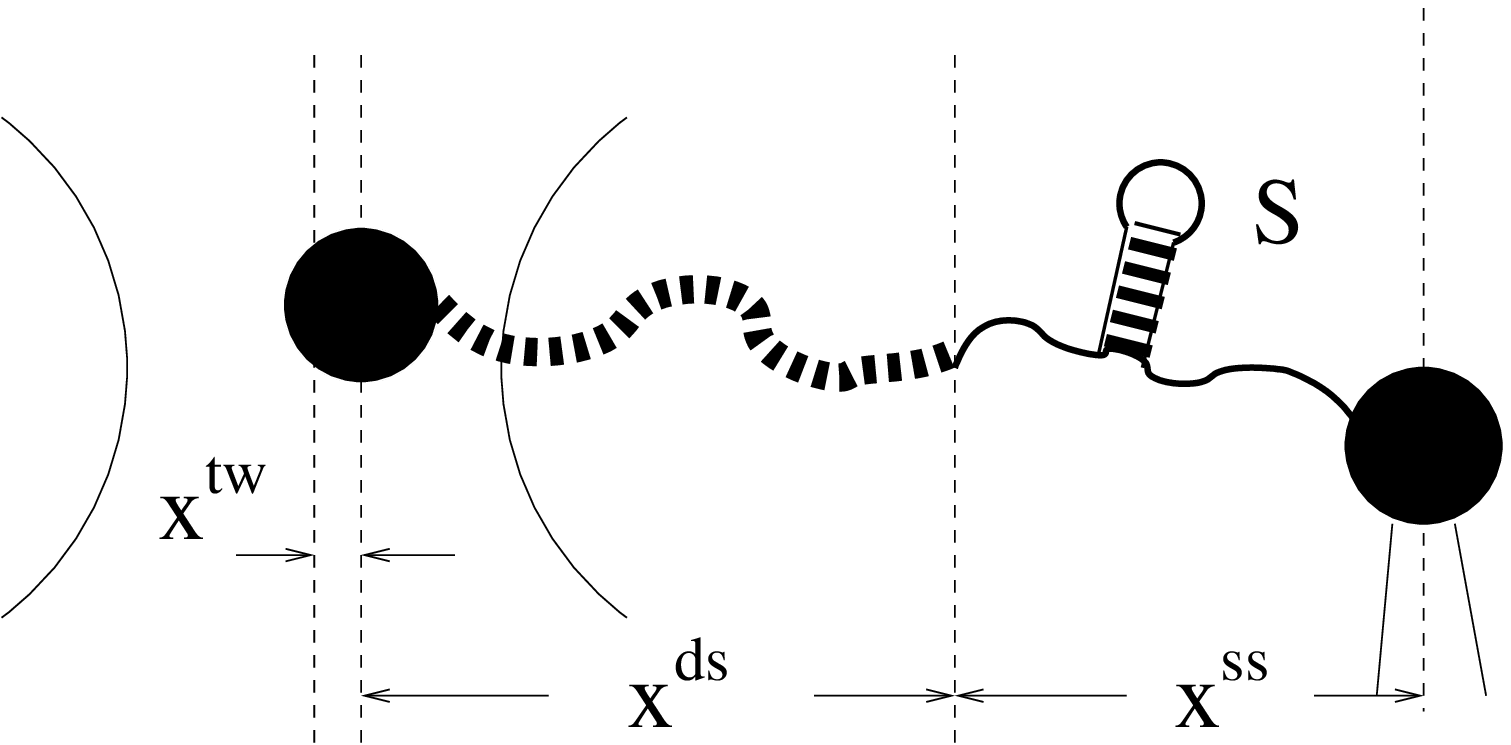}
\caption{Theoretical model and RNA sequences and their native
structures studied in present work. The structures are folded by
Vienna RNA package 1.4. The equilibrium and kinetic behaviors of
these three RNAs, p5ab, p5abc$\Delta$A, and p5abc have been
studied in detail \cite{liphardt}. } \label{experimentalsetup}
\label{figure1}
\end{center}
\end{figure}

According to the difference of the external
controlled parameters, the RNA unfolding experiments can be
carried out under constant extension and constant force, i.e., the
constant extension and the constant force ensembles
\cite{liphardt}. one of apparatus for the constant extension
ensemble is sketched in Fig.~\ref{figure1}: a single RNA molecule
is attached between two beads with RNA:DNA hybrid (double-stranded
DNA or dsDNA) handle; one bead is held by a pipette, and the other
is in a laser light trap. In practice, although two identical
handles connect the RNA, only one handle is considered in order to
simplify our theoretical calculation; It should not change
following discussions. By moving the position of the pipette, the
distance between the two beads and the force acting on the bead in
the light trap can be measured with high resolution. On the
contrary, constant force can be imposed on the RNA molecules with
feedback-stabilized optical tweezers capable of maintaining a
preset force by moving the beads closer or further apart.

\subsection{Monte Carlo simulation methods}
The Monte Carlo algorithm is simply reviewed in this section.

First is the method for the constant extension ensemble
\cite{liuf2}. Two simplifications have been made in our model. We
suppose that changes of the extensions of RNA and the handle
proceed along one direction. Physical effects of the beads are
neglected. Consequently, any state of the system can be specified
with three independent quantities, the position of the bead with
respect to the center of the optical trap, $x^{tw}$, the
end-to-end distance of the handle, $x^{ds}$, and the RNA secondary
structure $S$, i.e. the system in $i$-state
$(S_i,x^{tw}_i,x^{ds}_i)$. Here we do not include $x^{ss}$, the
extension of the RNA for the sum of individual extensions
satisfies constraint condition, $x=x^{tw}+x^{ds}+x^{ss}$, where
$x$ is the distance between the centers of the light trap and the
bead held by the pipet, and it also is the external controlled
parameter in the constant extension ensemble. The move set for
this system is as follow,
\begin{eqnarray}
&&(S_i,x^{tw}_i,x^{ds}_i)\rightarrow(S_j,x^{tw}_i,x^{ds}_i),
i\not=j\nonumber\\
&&(S_i,x^{tw}_i,x^{ds}_i)\rightarrow(S_i,x^{tw}_i\pm\delta,x^{ds}_i\mp\delta),\\
&&(S_i,x^{tw}_i,x^{ds}_i)\rightarrow(S_i,x^{tw}_i,x^{ds}_i\pm\delta).
\nonumber
\end{eqnarray}
Unfolding the single RNA for the constant extension ensemble
proceeds in an extended conformational space $C(l)\times
R^{tw}\times R^{ds}$, where $C(l)$ is the RNA secondary structural
folding space, $R^{tw}=(0,+\infty)$, $R^{ds}=(0,l_{ds})$, and
$l_{ds}$ is the contour length of the dsDNA handle. Given the
system state $i$, its whole energy is
\begin{eqnarray}
E_i(x)=\Delta
G^0_i+W^{tw}(x^{tw}_i)+W^{ds}(x^{ds}_i)+W^{ss}(x^{ss}_i,n_i),
\label{energyx}
\end{eqnarray}
where $\Delta G^0_i$ is the free energy obtained from folding the
RNA sequence into the secondary structure $S_i$, and the elastic
energies of the optical trap, the handle, and the single-stranded
part of the RNA are
\begin{eqnarray}
W^{tw}(x^{tw}_i)&=&\frac{1}{2}k_{tw}{x^{tw}_i}^2,  \nonumber\\
W^{ds}(x^{ds}_i)&=&\int_0^{x^{ds}_i}f_{ds}(x^\prime)dx^\prime,\\
W^{ss}(x^{ss}_i,n_i)&=&x^{ss}_if(x^{ss}_i,n_i)-\int_0^{f(x^{ss}_i,n_i)}x_{ss}(f^\prime,n_i)
df^\prime, \nonumber
\end{eqnarray}
respectively. In the expression $W^{ds}$, $f_{ds}(x^\prime)$ is
the average force of the handle at given extension $x^\prime$,
\begin{eqnarray}
f_{ds}(x^\prime)=\frac{k_BT}{P_{ds}}\left(
\frac{1}{4(1-x^\prime/l_{ds})^2}-\frac{1}{4}+\frac{x^\prime}{l_{ds}}
\right),
\end{eqnarray}
where $P_{ds}$ is the persistence length. respectively. In the
expression $W^{ss}$, $x_{ss}(f^\prime,n_i)$ is the average
extension of the single stranded part of the RNA whose bases
(exterior bases) is $n_i$ at given force $f^\prime$,
\begin{eqnarray}
x_{ss}(f^\prime,n_i)=n_ib_{ss}[\coth(\frac{f^\prime
l_{ss}}{k_BT})-\frac{k_BT}{f^\prime l_{ss}}],
\end{eqnarray}
where $b_{ss}$ and $l_{ss}$ are the monomer distance and the Kuhn
length of the single-stranded RNA, respectively
\cite{Bustamante,Smith}. Note that $f(x^{ss}_i,n_i)$ is the
inverse function of $x_{ss}(f^\prime,n_i)$.

Then is the simulation method for the constant force ensemble
\cite{liuf1}. We proposed an energy expression on the coarse-grain
level for the given secondary structure $S_i$ under constant force
$f$,
\begin{eqnarray}
E_i(f)\approx\Delta G^0_i - n_i \times g(f), \label{energyf}
\end{eqnarray}
where $g(f)=k_BTb_{ss}/l_{ss}\ln \sinh(u)/u$ and $u=l_{ss}f/k_BT$.
In contrast to the constant extension ensemble, the RNA secondary
structure $S$ can completely specify any state of the constant
force ensemble. Therefore, the move set for this ensemble is the
same with the set for RNA folding without force, i.e., its
unfolding space is $C(l)$.

Given the move sets and the unfolding conformational spaces, the
RNA unfolding for the two ensembles can be modelled as a Markov
process in their respective spaces. Define the transition
probabilities $k_{ij}$ from i-state to j-state satisfying
$k_{ij}=\tau_o^{-1}\exp(-\Delta E_{ij}/2k_BT)$, or the symmetric
rule \cite{Kawasaki}, where $\Delta E_{ij}=E_j-E_i$, $\tau_o$ is
used to scale time axis of the unfolding process. We use a
continuous time Monte Carlo algorithm to simulate unfolding
process \cite{BKL,Gillespie}.

The measurement quantities $\langle A \rangle$ for the two
ensembles can be calculated by $\langle A\rangle (n)=\sum_i
A_i(t_{i+1}-t_{i})$, where $A_i$ is the $A$-value in state $i$,
and $t_i$ is the inner time of the Monte Carlo simulations. For
the constant extension ensemble, $A$ could be the force exserted
on the bead, $f=k_{tw}x^{tw}$ in the light trap, or the
bead-to-bead distance $x^{bb}=x^{ds}+x^{ss}$. While for the
constant force ensemble, $A$ is the molecular extension $x$ under
the constant force $f$, and $x_i=x_{ss}(f,n_i)$. The simulation
time is $2\times 10^6\tau_o$.

\subsection{the exact numerical methods}
Compared to difficult protein folding prediction, the RNA
secondary structure prediction has achieved great success
\cite{Zuker}. In particular the partition function method
developed later provided strongly physical foundation
\cite{McCaskill}. Recently, this method was generalized to the
case of RNA unfolding in the constant extension ensemble
\cite{gerland}. In the present work, we are not ready to choose
the formulae presented in Ref.~\cite{gerland}. In addition to be
consistent with the formulae for the Monte Carlo simulations, the
complicated polymer model of single-stranded DNA (ssDNA) therein
might not result in many advances in predicting and understanding
the RNA unfolding phenomena.

The key idea of the partition function method is that the
partition function over all secondary structures of a given RNA
can be calculated by dynamic programming. Given the partition
function $Q(i,j,n)$ on the sequence segment [i,j] with exterior
bases $n$, its recursion formula is as follows,
\begin{eqnarray}
Q(i,j,n)&=&{\bf 1} \delta_{k,j-i+1}+qb(i,\Delta+j-n) \nonumber\\
&&+\sum_{k=i}^{j-1}\sum_{m=1}^{k-i+1}
Q(i,k,m)\\
&&\times qb(k+1,m+\Delta+j-n),\nonumber
\end{eqnarray}
where the partition function $qb(i,j)$ on the sequence segment
[i,j] for which the $i$ and $j$ bases are paired; Vienna package
1.4 provides their calculation codes \cite{Hofacker}.

For a given RNA sequence consisting of $N$ nucleotides, define
the total partition function for the constant extension and the
constant force ensemble $Z_N(x)$ and $Z_N(f)$, respectively.
According the energies mentioned in last section, their
expressions can be written as
\begin{eqnarray}
Z_N(x)&=&\sum_n^N \int_0^{l_{ds}} dx^{ds}\int_0^{nb_{ss}} dx^{ss} Q(1,N,n)\nonumber \\
&& \times \exp(-\beta E(x,x^{ds},x^{ss},n))
\end{eqnarray}
and
\begin{eqnarray}
Z_N(f)=\sum_n^N Q(1,N,n)\exp(-\beta E(f,n))
\end{eqnarray}
where the elastic energy
$E(x,x^{ds},x^{ss},n)=W^{tw}(x-x^{ds}-x^{ss})+W^{ds}(x^{ds})+W^{ss}(x^{ss},n)$
and $E(f,n)=n\times g(f)$. Correspondingly, the measurement
quantities for the constant extension ensemble are the average
force $\langle f\rangle=-k_BT\partial Z_N(x)/\partial x$ and the
average extension $\langle x^{bb}\rangle=x- \langle f\rangle
/k_{tw}$, and $\langle x\rangle=k_BT\partial Z_N(f)/\partial f$
for the constant extension ensemble, respectively.

\section{Comparison of the exact and simulation methods}
\label{result} To compare the exact and simulation methods
discussed above, we calculate extension-force curves of three
small RNA, p5ab, p5abc$\Delta A$ and p5abc in equilibrium. Their
native states under experimental condition are showed in Fig.~
\ref{figure1}. These molecules have been studied by the experiment
\cite{liphardt} and simulation \cite{liuf1,liuf2}. We first choose
the widely used parameters for our computation: temperature $T=298
K$, $b_{ss}=0.56$ nm, $l_{ss}=1.5$ nm, $P_{ds}=53$ nm
\cite{Smith,Bustamante}, $l_{ds}=320$ nm, and $k_{tw}=0.2$ pN/nm
\cite{liphardt}, and the free energy parameters for RNA secondary
structures at standard salt concentrations: $[Na^+]=1 M$ and
$[Mg^{2+}]=0 M$ \cite{Hofacker}.

\begin{figure}[htpb]
\begin{center}
\includegraphics[width=0.7\columnwidth]{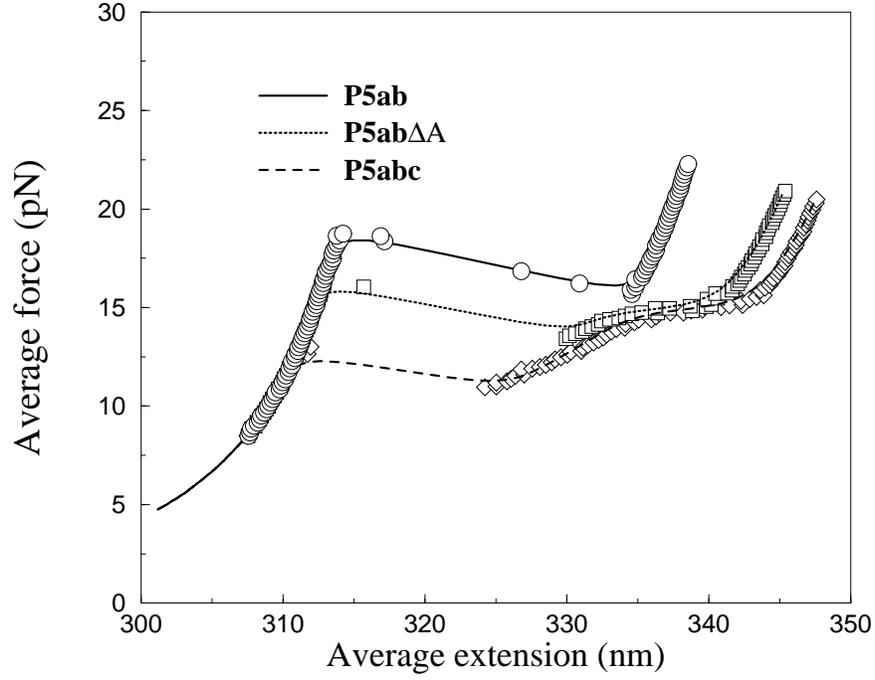}\\(a)\\
\includegraphics[width=0.7\columnwidth]{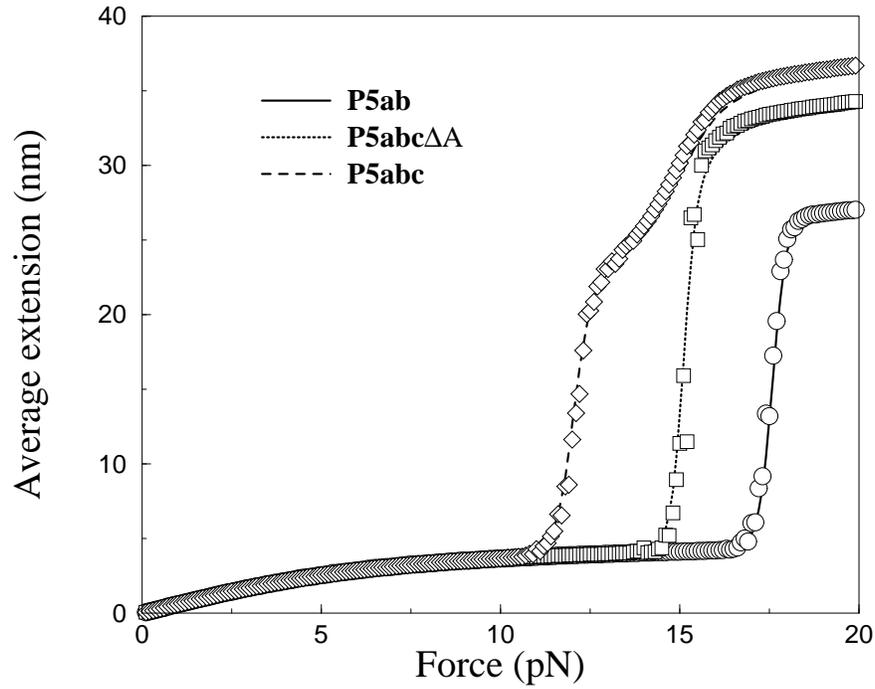}\\(b)
\caption{Comparison of the exact and simulation force-extension
curves in equilibrium for P5ab, P5abc$\Delta$A and P5abc in the
two ensembles. The different symbols are from the simulation
methods, and the different lines are from the exact methods. They
agree with each other very well.} \label{figure2}
\end{center}
\end{figure}

Fig.~ \ref{figure2} shows these extension-force curves for the
sequences for the two ensembles. We find that the two independent
methods achieve highly consistence. In particular, the three
curves of the molecules for the constant extension ensemble also
agree with the experimental measurements very well in quantity:
the extensional transition of P5ab are all-or-none; while P5abc
has an intermediate state \cite{liphardt}. Interestingly, we note
that, P5abc$\Delta A$ although has been observed as two-state
molecule in the experiment, a weaker intermediate state presents
in the constant extension ensemble, while it cannot be observed in
the constant force ensemble.

If we purchase the precision of our methods,  quantitative
comparison between the theoretical molecular unfolding force $f_u$
and the experimental measurements of course is essential. But we
find that they do not coincide: in the constant extension
ensemble, the experimental unfolding force of P5ab is $13.3$ pN,
of p5abc$\Delta$A is $11.4$ pN. and of P5abc is $8$ pN
\cite{liphardt}; while our calculations are $18.4$ pN, $15.8$ pN
and $12.2$ pN, respectively. So what causes result to the larger
differences between the experiment and the theory? The experiment
and previous theoretical works contributed the differences to the
change of free energy of the RNA secondary structure; this change
results from the different ionic concentration of the experiment
and the standard condition: in the RNA unfolding experiment,
$Na^+=250$ mM and with and without $Mg^{2+}=10$ mM
\cite{liphardt,gerland,cocco}. To reproduce the experimental ionic
condition, a correction on the energy of a base pair equal to
$-0.193k_BT\ln ([Na^+]+3.3[Mg^{2+}]^{1/2})$ has been applied
\cite{cocco}. Their values are summarized in Tab. \ref{table}.
Besides the three molecules from Ref. \cite{liphardt}, other
unfolding forces of the molecules published in the lectures
\cite{Bustamante_rev,rief} are also listed there. We still see
that the ionic correction cannot explain the derivation between
the theory and experiment.

Considering that the free energy parameters of the RNA secondary
structure were measured in bulk experiments, one might doubt
whether they can be used in single-molecule studies as well as we
thought before \cite{liuf2}. On the other hand, however, it is
known that the mechanical parameter, the persistent length
$l_{ss}$ is also sensitive to ionic condition. Although this
parameter indeed were measured under a similar experimental
conditions with the small RNA unfolding experiment (see Ref.
\cite{Smith}), their validity for describing small molecules is
questionable. Recent FRET experiment measured that for shorter
ssDNA $l_{ss}$ is about 2.2 nm at $Na^+=250$ mM \cite{Murphy}. If
we choose this value in our calculation, the predicted unfolding
forces are closer with the experimental measurements; see Tab.
\ref{table}. Of course, we cannot exclude the intrinsic limitation
of our coarse-grain model. For example, another possible force
work formula has been used in the constant force ensemble
\cite{liuf1}.

\section{conclusion}\label{conclusion}
In this work, we review the Monte Carlo methods and develop the
exact numerical methods to study the force stretching single RNA
molecules issue. We respectively compare the two independent
method in the constant force and extension ensembles, and find
that they agree with each other quite well. We also point out that
only ionic correction on the RNA secondary structure alone cannot
explain the larger discrepancies of the unfolding forces between
the theoretical prediction and the experimental measurement; the
ionic correction on the RNA molecular mechanical properties should
be important.

Although the results of the exact numerical method are consistent
with the Monte Carlo method when force stretches single RNA in
equilibrium, it does not mean the former can completely replace
the later. Such situation is similar with the study of 2-dimension
Ising model in condense matter physics \cite{Newman}. Compared to
the exact method, the Monte Carlo method would be more
sophisticate in dealing possible more complicated experimental
condition. For instance, recent simulation work could include
pseduknots structure \cite{Isambert}, while the exact partition
function technique would be hardly to realize. In our point of
view, Monte Carlo simulation is more important in studying single
molecular non-equilibrium behavior produced by mechanical force,
such as folding/unfolding trajectories, force-hysteresis phenomena
and unfolding force dependance on loading rates etc
\cite{liuf1,liuf2}.

We thank Professor H.-W. Peng for many helpful discussions in this
work.

\newpage
\begin{table*} \caption{The unfolding forces $f_u$ of
different molecules under different experimental conditions. The
experimental data are from the previously published data
\cite{liphardt,Bustamante_rev,rief}. The theoretical values are
from the exact numerical methods developed above, where $f^i_u$,
$i=1,2,3$ represent the unfolding forces without the ionic
correction, with the ionic correction on the free energy and with
the ionic and the persistent length corrections, respectively.
Here We do not show the P5abc unfolding force for it is not
reversible in $Mg^{2+}$ due to the presence of tertiary
interactions.}
\begin{center}
\begin{tabular}{cccccccc}
\hline\hline Molecule& temperature (K) &  $Na^+$ (mM) & $Mg^{2+}$
(mM)& $f^1_u$ (pN)& $f^2_u$ (pN)& $f^3_u$ (pN) &
$f^{exp}_u$ (pN)\\
\hline
P5abc&298&250&0&12.2&11.4&10.0&7.0-11.0\\
poly(dA-dU)&293&150&0&12.3&11.0&9.3&9.0 \\
P5abc$\Delta$A&298&250&0&15.8&14.8&13.2&$11.4\pm0.5$\\
P5abc$\Delta$A&298&250&10&&15.4&13.8&$12.7\pm0.3$\\
P5ab &298&250&0&18.4&17.4&15.7&$13.3\pm1.0$\\
P5ab &298&250&10&&18.0&16.2&$14.5\pm1.0$\\
CG hairpin&293&150&0&25.8&24.4&22.4&17.0\\
poly(dC-dG)&293&150&0&25.1&23.8&21.7&20.0\\
\hline\hline
\end{tabular}
\label{table}
\end{center}
\end{table*}

\end{document}